\documentclass[english]{emulateapj}
\usepackage[T1]{fontenc}
\usepackage[latin1]{inputenc}
\usepackage{graphicx}
\usepackage{amssymb}
\usepackage{amsmath}
\usepackage[usenames]{color}
\usepackage{babel}
\usepackage{natbib}

\definecolor{DarkBlue}{rgb}{0.1,0,0.6}

\newcommand{\msun}{M_\odot}
\newcommand{\mbh}{M_\bullet}
\newcommand{\nob}{N_\mathrm{OB}}
\newcommand{\rhs}{r_\mathrm{h}}
\newcommand{\pc}{\mathrm{pc}}
\newcommand{\myr}{\mathrm{Myr}}
\newcommand{\rd}{\mathrm{d}}
\newcommand{\au}{\mathrm{AU}}

\newcommand{\kms}{\mathrm{km\,s}^{-1}}

\newcommand{\rlim}{3}

\bibpunct{\hspace{-4pt}}{}{}{a}{}{} 

\shorttitle{Catch me if you can}
\shortauthors{\v{S}ubr et al.}

\begin{document}
\title{Catch me if you can:  is there a `runaway-mass' black hole\\
in the Orion Nebula Cluster?}

\author{Ladislav \v{S}ubr\altaffilmark{$1$},
Pavel Kroupa\altaffilmark{$2$}
and Holger Baumgardt\altaffilmark{$3$}}

\email{subr@mbox.troja.mff.cuni.cz}

\altaffiltext{$1$}{Astronomical Institute, Charles University,
V Hole\v{s}ovi\v{c}k\'ach 2, CZ-180\,00 Praha, Czech Republic}
\altaffiltext{$2$}{Argelander Institute for Astronomy (AIfA), Auf
dem H\"ugel 71, D-53121 Bonn, Germany}
\altaffiltext{$3$}{University of Queensland, School of Mathematics and Physics, Brisbane, QLD 4072, Australia}

\begin{abstract}
We investigate the dynamical evolution of the Orion Nebula Cluster (ONC) by
means of direct $N$-body integrations. A large fraction of residual gas was
probably expelled when the ONC formed, so we assume that the ONC was much more
compact when it formed compared to its current size, in agreement with the
embedded cluster radius-mass relation from \cite{marks12}.
Hence, we assume that few-body relaxation played an important role during the
initial phase of evolution of the ONC. In particular, three body interactions
among OB stars likely led to their ejection from the cluster and, at the same
time, to the formation of a massive object via `runaway' physical stellar
collisions. The resulting depletion of the high mass end of the stellar mass
function in the cluster is one of the important points where our models fit the
observational data. We speculate that the runaway-mass star may have
collapsed directly into a massive black hole ($\mbh \gtrsim
100\msun$). Such a dark object could explain the large velocity dispersion of
the four Trapezium stars observed in the ONC core.
We further show that the putative massive black hole is likely to be a member
of a binary system with $\approx70\%$ probability. In such a case, it could be
detected either due to short periods of enhanced accretion of stellar winds
from the secondary star during pericentre passages, or through a measurement
of the motion of the secondary whose velocity would exceed
$10\,\mathrm{km\,s}^{-1}$ along the whole orbit.
\end{abstract}

\section{Introduction}
The Orion Nebula Cluster (ONC, M42) is a dense star cluster which is part
of a complex star forming region at a distance of about $400 \pc$
\citep{jeffries07,sandstrom07,menten07}. Due to its relative proximity the ONC
is one of the best observationally studied star clusters. Its age is estimated
to be $\lesssim 3\myr$, the resolved stellar mass is
$M_\mathrm{c} \approx 1800\msun$ and it has a compact core of radius
$\lesssim 0.5\pc$ \citep{hh98}. Based on the data presented by \cite{hs06},
we estimate a half-mass radius of the cluster of $\rhs \approx 0.8\pc$. Hence,
the ONC is considered to be a prototype of a dense young star cluster and
it naturally serves as a test bed for theoretical models of various
astrophysical processes (e.g. \citealt{kroupa01} considered the dynamical
evolution of ONC-type star clusters with gas expulsion; \citealt{olczak08}
studied the destruction of protoplanetary discs due to close stellar encounters).

Nevertheless, the morphological and dynamical state of the ONC is still not
fully understood. Some of it's characteristics indicate that it has
undergone a period of violent evolution and that the current state does not
represent a true picture of a newly born star cluster. There is a lack of
gas in the cluster \citep{wilson97} which has been expelled due to the radiation
pressure of the OB stars that reside in the cluster core. The effect of gas
removal is likely to have led to the cluster expanding by a factor $\gtrsim3$
for a star formation efficiency $\lesssim50\%$ \citep[e.g.][]{bk07}.
The assumption that the ONC has been more compact in the past is in accord with
the fact that this star cluster is missing wide binaries with separations
$>10^3 \au$ which may have been disrupted through close three-body interactions
during the initial compact stage \citep{kroupa00, parker09, marks12}.
The process of dynamical ejections has
been suggested by \cite{pflamm} as a reason for the depletion of the cluster mass
function at the high-mass end. This hypothesis aims to explain the observed deficit
of massive stars in the ONC \citep{hillenbrand97} with respect to the standard
\cite{kroupa01b} initial mass function. Finally, observations \citep{zapata09}
have brought evidence that stellar disruptions occurred in the ONC recently.
These may be a consequence of physical stellar collisions in
the dense cluster core.

In this paper we address the history of
the ONC over its lifetime. We concentrate on initialy
very compact star clusters and show that they can evolve within a few
millions years into a state compatible with the current observations.
In particular, we address the fact that the central system of OB stars, the
so-called Trapezium, is supervirial and, at the same time, that the ONC hosts
unexpectedly few OB stars.

\section{Model}
We restricted ourselves to the stage of cluster evolution
when the stars have already formed as individual entities which can be
characterised by a constant mass and radius. Stellar dynamics is then driven
mainly by gravitational interactions. We used the numerical code NBODY6
\citep{aarseth03} which is a suitable tool for modeling
self-gravitating stellar systems with a considerable amount of binaries.

At the initial stage of its evolution, a substantial contribution to the
gravitational field of the cluster is due to gas. We incorporated
external gas into our model by means of a special type of low mass particles
($m_\mathrm{g} \leq 0.4\,\msun$) whose gravitational interaction with the rest
of the cluster was treated directly via the N-body scheme. Beside that, we
modified the original NBODY6 code, including an option for a {\em repulsive\/}
force (with the direction radial from the cluster centre) acting upon the gas
particles in order to mimic the radiation pressure from the stars starting
at a given time $T_\mathrm{ex}$ (see
Sec.~\ref{sec:expulsion}). Alternatively, we also modelled the gas
expulsion by an instantaneous removal of the gas particles at a given time.

Stellar masses were calculated according to the \cite{kroupa01b}
mass function with an upper mass limit of $80\,\msun$ (however, the mass of
the most massive star is typically $\approx63\,\msun$). For several
numerical reasons (e.g. quadratic growth of CPU time with the number of
particles, higher probability of numerical errors for extreme mass ratios)
we replaced low-mass stars ($M_\star \leq m_\mathrm{min}$) by stars with mass
$m_\mathrm{min}$, keeping the total mass unchanged. We have verified (cf.
models \#4 and \#6 introduced below which have $m_\mathrm{min} = 0.5\,\msun$ and
$0.2\,\msun$, respectively) that our results are not affected considerably by
this approximation.

The stars were considered to have finite radii $R_\star = R_\odot(M_\star
/ \msun)^{0.8}$ (e.g. \citealt{lang}; this simple formula also fits well
the radii of zero age main sequence stars for $M_\star \gtrsim \msun$ according
to \citealt{eggleton89}; our overestimation of stellar radii for
$M_\star \lesssim \msun$
does not affect our results, as the low-mass stars contribute only marginally
to the merging tree). If the separation of two stars got below the sum
of their radii, they were merged into a single star. We also switched off the
option of stellar evolution in the numerical code. Loss of realism is assumed
to be negligible as we followed the cluster evolution for a period of only
$2.5\myr$. The stars in our models are not assigned a
spectral type. For the sake of brevity, we refer to all stars with
$M_\star \geq 5\,\msun$ as `OB stars'.

\begin{figure}[t]
\begin{center}
\includegraphics[width=\columnwidth]{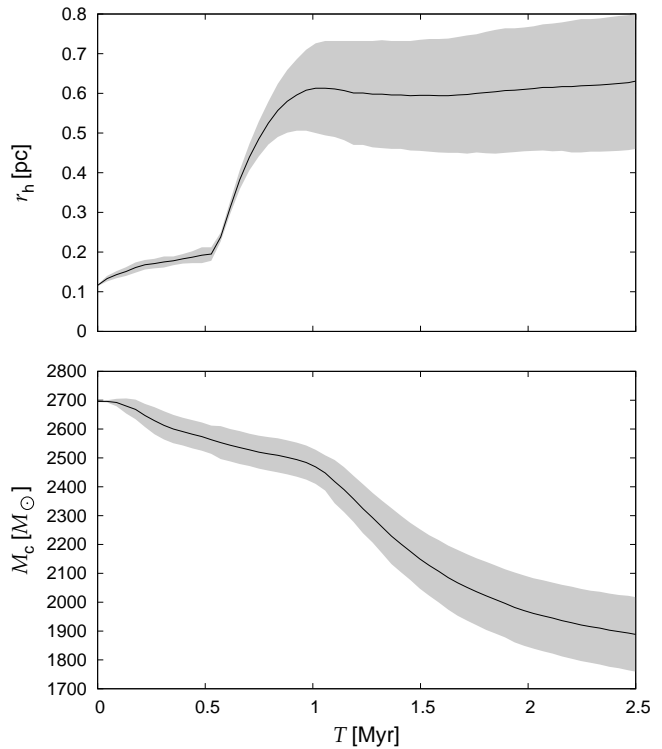}
\end{center}
\caption{Half-mass radius of the star cluster of our canonical model of the
ONC (upper panel). Expansion is accelerated at $T\approx0.5\myr$
due to gas expulsion. The total mass of the stars within a $\rlim\pc$ radius
(lower panel) monotonically decreases as stars escape out of the
observer's field of view. This curve reflects gas expulsion with a time delay
of $\approx0.5\myr$ which is the time the stars need to reach the $\rlim\pc$
boundary. Shaded area represents the $1\sigma$ variance of the individual
realisations of the model.}
\label{fig:rh}
\end{figure}

\subsection{The `canonical' model}
The canonical (best-fit) model of the ONC has an initial mass of
$5400\msun$, half of which is in the form of stars, while the other half
accounts for gas. We considered a \cite{kroupa01b} initial mass function
which, for the given cluster star mass, predicts $\approx 50$ OB stars to have
been formed in the ONC \citep{pflamm}. In order to avoid additional statistical
noise in the evolution of the number of OB stars, we used identical samples of
stellar masses for all numerical realisations of the model. The initial
half-mass radius of the cluster is $r_\mathrm{h} \approx 0.11\pc$, which is
compatible with the initial or birth radius-mass relation inferred for star
clusters ranging in mass up to globular clusters \citep{marks12}.
Positions and velocities are generated in a mass segregated state
according to the prescription of \citet{skh08} with mass segregation index
$S = 0.4$. The algorithm by definition places massive stars in the cluster
core and, therefore, the initial half-mass radius of the OB stars is
$\approx 0.05\pc$. In order {\em not\/} to place the light gas particles at
the cluster outskirts, we generated initial positions for stellar and gas
particles separately, i.e. we had a constant gas to star mass ratio throughout
the whole cluster at $T=0$. All OB stars were set to be members of primordial
binaries with a secondary mass $M_\mathrm{s} \geq 1\msun$ with the pairing
algorithm being biased towards assigning a massive secondary to a massive
primary. More specifically, the algorithm first sorts the stars from the
most massive to the lighest one. The most massive star from the set is taken as
the primary. The secondary star index, $id$, in the ordered set is generated as a
random number with the probability density $\propto id^{-\beta}$ and $\max(id)$
corresponding to a certain mass limit, $M_\mathrm{s,min}$. The two stars are
removed from the set and the whole procedure is repeated until stars with
$M_\star \geq M_\mathrm{p,min}$ remain. Typically, we used as the minimal mass
of the primary $M_\mathrm{p,min} = 5\msun$, as the minimal mass of the secondary
$M_\mathrm{s,min} = 1\msun$ and the pairing algorithm index $\beta=40$.
For the assigned binaries we used a semi-major axis distribution according
to \"Opik's law $n(a)\propto a^{-1}$ \citep{kobulnicky07} and a thermal
distribution of eccentricities, $n(e)\propto e$. Except for the possibility of
being a secondary member of a binary containing an OB star, we did not generate
binaries of low mass stars. This limitation comes from the requirement of
numerical effectivity and it is not likely to affect our results substantially.

Let us note that the canonical model presented above
fits into the range of the expanding class of models of the ONC discussed by
\cite{kroupa00}. Other possible initial conditions would be fractal models
that collapse \citep[e.g.][]{allison09}. These require star formation to be
synchronised across the pre-cluster cloud core to much shorter than the dynamical
time though.

\subsection{Dynamical evolution}
Most of the features of the evolution of a star cluster with strong
few-body relaxation and gas expulsion can be demonstrated with the canonical
model. In order to distinguish systematic effects from rather large
fluctuations of cluster parameters, we present quantities averaged over $100$
realisations of the model. The upper panel of Fig.~\ref{fig:rh}
shows the temporal evolution of the half-mass radius of the stellar component,
$\rhs$. During the initial phase, the cluster expands due to two-body
relaxation. Gas expulsion that starts at $T_\mathrm{ex} = 0.5\myr$ removes
all gas from the cluster within a few hundred thousand years. From the point
of view of the stars, this event leads to an abrupt decrease of their potential
energy. Consequently, they can reach larger distances from the cluster centre,
which manifests itself as an accelerated growth of $\rhs$ at
$0.5\myr \lesssim T \lesssim 1\myr$. Initially weakly bound stars became
unbound due to the gas expulsion and they escape from the cluster. Hence,
reduction of the cluster mass also accelerates -- see the lower panel of
Fig.~\ref{fig:rh} where we plot the mass of the star cluster. We define
$M_\mathrm{c}$ as the sum of the mass of the stars within a sphere of radius
$r_\mathrm{lim}$ from the cluster centre. We set $r_\mathrm{lim} = \rlim\,\pc$
in order to match typical observations of the ONC, which usually count
stars within a projected distance of $\approx \rlim\pc$ from the Trapezium.
Note that the decrease of $M_\mathrm{c}$ is accelerated with $\approx 0.5\myr$
delay after the time of gas expulsion. This is due to the unbound stars taking
some time to reach $r_\mathrm{lim}$. Besides the escape of stars due to the
gas expulsion, there is also continuous stellar mass loss due to
the few-body relaxation which is capable to accelerate stars above the escape
velocity. From $T\approx1\myr$ onwards the cluster expansion is again dominated
by relaxational processes driving a revirialisation \citep{kroupa01}.
\begin{figure}[t]
\begin{center}
\includegraphics[width=\columnwidth]{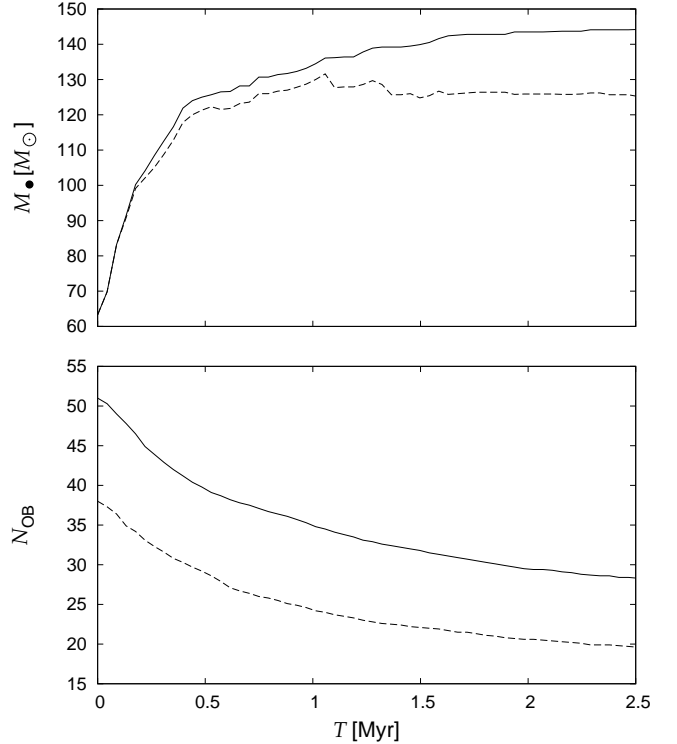}
\end{center}
\caption{Top: Mean mass of the most massive (collisional) star in the model
regardless of its location (solid line) and mean mass of the most massive star
within $\rlim\pc$ from the cluster centre (dashed line).
Bottom: number of OB stars (solid line) and number of OB systems (dashed
line) within $\rlim\pc$ from the cluster centre.}
\label{fig:nob}
\end{figure}

Our model indicates that the initial mass of the ONC must have been at least
20\% larger than it is now, after a few million years of dynamical evolution.
We also see that, mainly due to the effect of gas expulsion, the ONC must
have been more compact by a factor $\gtrsim 5$ at the time of its birth than
it is now. Hence, the initial relaxation time was much shorter than what we
would infer from present-day observations. Consequently, close few-body
interactions must have been more frequent in the past.
In particular, we suggest that scattering of single stars on binaries is
a process that has played an important role in the evolution of the ONC.
This kind of interaction has been studied intensively in the literature
(e.g. \citealt{hh} and references therein). A rich variety of outcomes can be
obtained, depending on the initial conditions. Nevertheless, some general
results can be formulated. In particular, if the binary is hard, i.e. its
orbital velocity is larger than the impact velocity of the third star,
it typically shrinks, losing its energy which is transferred to the
accelerated single star. Therefore, we expect two
processes to happen in correlation: (i) ejections of high velocity stars
and (ii) physical collisions of massive binaries which lead to the formation
of more massive objects.

The probability of collisions increases with the mass of the interacting stars.
Therefore, numerical models by different authors often
show a `runaway' process when most of the collisions involve the most massive
object which grows continuously \citep[e.g.][]{spz04}. The same is true also
for our model of the ONC: The upper panel of Fig.~\ref{fig:nob} shows the
growth of the mass of the most massive object, $\mbh$, due to stellar
collisions; the number of OB stars left in
the cluster is plotted in the lower panel. Approximately one third of the
missing OB stars have disappeared due to merging,
while the other two thirds have been ejected from the cluster with velocities
$>10\mathrm{km\,s}^{-1}$. The merging tree varies significanly among individual
realisations of the particular model. In general, more than one merging star
is formed during the initial $\approx0.5\myr$. Among other processes, this is
due to the merging of several primordial OB binaries. At later stages, usually
one runaway-mass object dominates the merging process.

In the Appendix we provide an approximate description of the rate of stellar
ejections due to the scattering of single stars on massive binaries. It gives
an estimate of the decay time of the number of the OB
stars of $\tau\gtrsim5\myr$ for the canonical model. If we assume that the decay
rate is linearly proportional to the number of OB stars, we
expect roughly an exponential decay of $\nob$, i.e. it should reach half of
its initial value at $\sim3.5\myr$,  which is in good
agreement with the numerical results.

\begin{figure}[t]
\begin{center}
\includegraphics[width=\columnwidth]{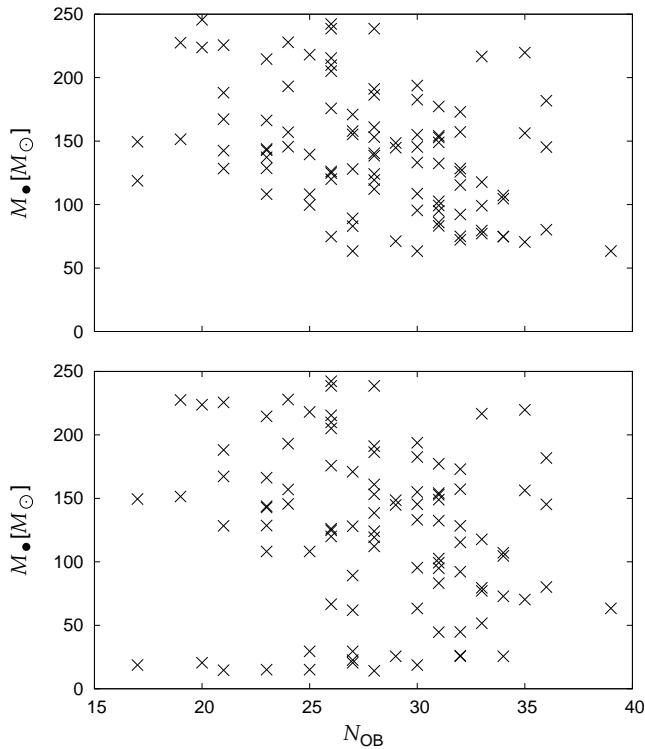}
\end{center}
\caption{Mass of the most massive object in the cluster vs. the number of OB
stars within a radius of $\rlim\pc$. Individual points represent states of
different realisations of the canonical model at $T = 2.5\myr$. Mass of the
most massive object in the whole set is plotted in the upper panel, while
in the lower one, only objects found within $\rlim\pc$ are considered.}
\label{fig:mn}
\end{figure}
A mean number of OB stars of $\approx28$ remain at $T=2.5\myr$. This considerably
exceeds the number of OB systems observed in the ONC --- the study of
\cite{hillenbrand97} reports only 10 stars heavier than $5\msun$. However, the
real number of massive stars may be somewhat larger due to their `hiding' in OB
binary systems. Our models always end up with several OB stars having
an OB companion and, therefore, a better agreement with the observations is
achieved if we count the OB {\em systems\/}
as indicated with the dashed line in Fig.~\ref{fig:nob}. Moreover, due to
the stochastic nature of dynamical cluster evolution,
different realisations of the model end up with quite different values of
$\nob$ and $\mbh$. The final state ($T=2.5\myr$) of all realisations of the
canonical model is shown in Fig.~\ref{fig:mn}. Although the distribution
of points in the $\nob$--$\mbh$ space is rather noisy, we can deduce an
anti-corelation between these two quantities (their linear correlation
coefficient is $-0.42$). In particular, {\em all\/} realisations that end up
with $\nob\leq20$ lead to the formation of a runaway-mass star of mass
$\mbh > 100\msun$. In other words, the underabundance of the high-mass stars
in the ONC not only indicates a period of prominent two-body relaxation in the
past, but also the merging formation of a massive object
which may represent an important footprint of the cluster's history.

Interestingly, the term `runaway' has a secondary meaning in some cases:
Despite of its high mass, in several realisations the merging object has been
ejected out of the cluster with velocity exceeding $10\,\kms$. These cases can
be identified due to differences of
the upper and lower panels of Fig.~\ref{fig:mn}. The escape of the most massive
body beyond the $\rlim\pc$ boundary is also responsible for the drops of the
dashed line in the plot of the temporal evolution of $\mbh$ in
Fig.~\ref{fig:nob} which shows the mean mass of the most massive object within
$\rlim\pc$ from the cluster centre.

\section{Discussion}
\begin{table}[t]
\begin{center}
\caption{Initial parameters and final states of several models.}
\label{tab:models}
\begin{tabular}{cccc|cccc}
id & $\!\! S \!\!$ & $\!\!\! T_\mathrm{ex}/\myr \!\!\!$ & 
 $\!\! \epsilon$ & $\!\rhs/\pc \!\!$ & $\!\! M_\mathrm{c}/\msun \!\!$ & 
 $\!\! N_\mathrm{OB} \!\!$ & $\!\! \mbh/\msun \!\!$ \\
\hline
1 & 0.00 & 0.5 & 0.50c & 0.56 & 2183 & 35.1 & 102.9 \\ 
2 & 0.00 & 0.7 & 0.50c & 0.48 & 2061 & 28.1 & 138.1 \\ 
3 & 0.25 & 0.7 & 0.50c & 0.51 & 1980 & 26.2 & 140.9 \\ 
4 & 0.40 & 0.5 & 0.50c & 0.59 & 1895 & 28.0 & 140.1 \\ 
5 & 0.40 & 0.5 & 0.50c & 0.60 & 1938 & 27.3 & 138.5 \\ 
6 & 0.40 & 0.5 & 0.50c & 0.63 & 1887 & 28.3 & 144.2 \\ 
7 & 0.40 & 0.5 & 0.50c & 0.66 & 1660 & 26.0 & 142.7 \\ 
8 & 0.40 & 0.5 & 0.50v & 0.47 & 2197 & 29.7 & 122.1 \\ 
9 & 0.40 & 0.7 & 0.50c & 0.62 & 1895 & 26.5 & 152.1 \\ 
10 & 0.40 & 0.7 & 0.50c & 0.58 & 1882 & 27.0 & 144.2 \\ 
11 & 0.40 & 0.7 & 0.50c & 0.67 & 1636 & 26.1 & 132.6 \\ 
12 & 0.00 & 0.7 & 0.33c & 0.52 & 1717 & 24.2 & 173.7 \\ 
13 & 0.25 & 0.5 & 0.33v & 0.51 & 2109 & 32.1 & 124.2 \\ 
14 & 0.40 & 0.5 & 0.33v & 0.50 & 2196 & 33.1 & 112.1 \\ 
15 & 0.40 & 0.5 & 0.33c & 0.77 & 1300 & 23.2 & 171.0 \\ 
\end{tabular}
\end{center}
Common initial parameters of all models are the initial mass of the stellar
component, $M_\mathrm{c} = 2700\msun$ which implies $\nob(T=0) = 50$ and
$\mbh(T=0) \approx 63\msun$. The initial half-mass radius varies slightly
throughout the models, but is generally $\approx 0.1\pc$. The gas particles are
of mass
$0.2\msun$ for models 6 and 10; in all other cases $m_\mathrm{gas} = 0.4\msun$
is considered. In models 7 and 11, gas particles were removed instantaneously
at $T=T_\mathrm{ex}$; in all other cases, $T_\mathrm{ex}$ is the time when
the repulsive force was switched on. $S$ is the mass segregation index as
defined in \cite{skh08}. $\epsilon\equiv M_\mathrm{c} / (M_\mathrm{c} +
M_\mathrm{gas})$ is the star formation efficiency; suffix `c' stands
for constant $\epsilon$ throughout the whole cluster, while `v' means variable
$\epsilon$ (decreasing outwards). \"Opik's distribution of semi-major axes of the
primordial binaries is considered in all models except for \#5, where
we set $n(a)=\mathrm{const.}$ The canonical model has id 6.
\end{table}
The final state of the canonical model presented in the previous section matches
basic observables of the ONC quite well, in particular its mass and half-mass
radius. The mean number of OB stars is somewhat larger than what is observed
in the ONC, nevertheless, some of the realisations reach $\nob < 20$. Hence,
we consider this model to be a realistic representation of the ONC. We have
investigated several tens of different models with different values of the
parameters including the index of the initial mass segregation, initial
half-mass radius, mass of the cluster, star to gas mass fraction and the time
of the gas expulsion. Models whose final state can be considered
compatible with the current state of the ONC, at least in some of their
characteristics, are listed in Table~\ref{tab:models}.

\subsection{$\nob - \mbh$ anticorelation}
\begin{figure}[t]
\begin{center}
\includegraphics[width=\columnwidth]{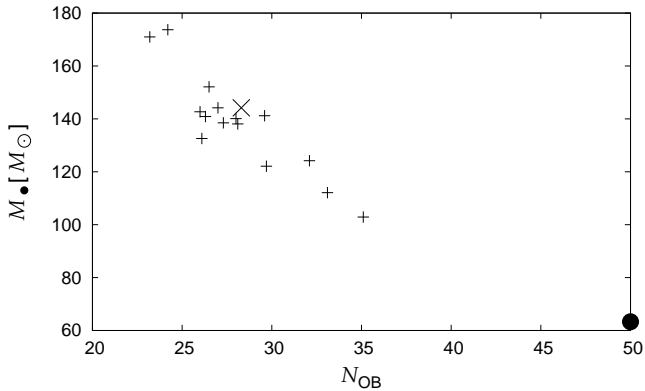}
\end{center}
\caption{Mean mass of the most massive object in the cluster vs. the mean
number of OB stars (both within $3\,\pc$) at $T=2.5\mathrm{Myr}$ for several
different models of
compact star clusters listed in Table~\ref{tab:models}. The canonical model
is represented with the $\times$ sign; full circle stands for
the initial state which is common for all models.}
\label{fig:mn_avg}
\end{figure}

As mentioned above, the canonical model indicates an anticorelation between
the final number of the OB stars and the final mass of the runaway-mass
star. In Fig.~\ref{fig:mn_avg} we plot the mean mass of the most massive
object vs. the mean number of remaining OB stars for all models listed in
Table~\ref{tab:models}. Now, each point represents an average over several
tens of realisations of the particular model and a $\nob - \mbh$
anticorelation becomes evident.
We attribute this relation to the fact that both mechanisms of OB star
removal, i.e. physical collisions and ejections, are driven by a common
underlyig process of close three-body interactions
(see Appendix). Naturally, their importance grows with increasing stellar
density. Hence, the models that stay more compact during the course of their
evolution are located in the top-left corner of the graph.
They better fit the observations from the point of view of the number of
OB stars, however, either their half-mass radius or total stellar mass
at $T=2.5\myr$ is too small, i.e. not consistent with the observations.

\subsection{Gas expulsion}
\label{sec:expulsion}
In most of our models, the gas particles were blown out of the cluster due
to a repulsive external force $\propto \boldsymbol{r}/r^3$ centered on the
cluster core. The strength of the force was set such that the gas particles
were accelerated to velocities $\approx8 \mathrm{km\,s}^{-1}$ on the time-scale
of $\approx 0.1\,\myr$. This approach
is the most realistic one, keeping the continuity equation fulfilled, but, at
the same time, it is numerically the most expensive. The alternative method
with instantaneous removal of the gas particles at $T_\mathrm{ex}$ leads to
similar results. The clusters integrated with
this method typically result in somewhat lower total mass and larger half-mass
radius at the final time. Let us note, for completness, that in the literature,
another method that mimics gas removal in numerical models of star clusters is
also used \citep[e.g.][]{geyer01,kroupa01,bk11}. It is based on the
time-variable smooth external potential that represents the weakening
gravitational potential of the gas. It has been found that the two approaches
(i.e. modelling gas via particles vs. external potential) lead to nearly
indistinguishable results \citep{geyer01}. In order to check for a possible
bias of our gas particles approach which may stem from two-body relaxational
processes, we have integrated two additional models with very low masses for the
gas particles. Due to the large number of particles in these integrations, we have
followed their evolution only up to the time of the gas expulsion, $T_\mathrm{ex}
= 0.5 \,\myr$. Outcomes of these models in terms of the mean number of remaining
OB stars, $\nob$, and the mass of the runaway-mass star, $\mbh$, are presented in
Table~\ref{tab:mgas}. Besides just giving the mean values of $\nob$ and $\mbh$,
we have also performed a Kolmogorv-Smirnov test comparing the sets of $\nob$
and $\mbh$ coming from these models with the canonical one. As can be seen, there
is no apparent dependence of the results on $m_\mathrm{gas}$ ranging from $0.03$
to $0.4\,\msun$. Hence, we conclude that the two-body relaxation due to the gas
partiles  does notsignificantly affect our results.
\begin{table}
\begin{center}
\caption{Comparison of models with different masses of gas particles}
\label{tab:mgas}
\begin{tabular}{ccccccl}
$\!\! m_\mathrm{g} \!\!$ & $\!\! \nob \!\!$ & $\!\! \mbh \!\!$ & 
$\!\! N_\mathrm{run} \!\!$ & $\!\! P_\mathrm{KS,N_{OB}} \!\!$ &
$\!\! P_\mathrm{KS,\mbh} \!\!\!$ & comment \\
\hline \\
0.40 & 39.7 & 125.1 &  40 & 95\% & 42\% & model \#4 \\
0.20 & 39.8 & 124.9 & 100 &  --- &  --- & canonical model \\
0.05 & 38.2 & 128.5 &  14 & 82\% & 87\% & \\
0.03 & 39.9 & 126.7 &  14 & 32\% & 75\% &
\end{tabular}
\end{center}
Mean values of the number of OB stars within the radius of $3\,\pc$ from the
cluster core, $\nob$, and mass of the runaway-mass star, $\mbh$, for models with
different values of $m_\mathrm{g}$. All other parameters of the models, including
the total mass of gas, are identical to the canonical model. $P_\mathrm{KS,\ast}$
is the Kolmogorov-Smirnov probability that the null hypothesis (assuming the
particular data set comes from the same distribution as the canonical one)
is valid. Note that the value of $P_\mathrm{KS,\ast} \approx 5\%$ is usually
considered to be a limit below which the null hypothesis should be rejected.
\end{table}

The evolution of the radiation pressure in the real ONC has been definitely
more complicated. According to observations \citep{hs06}, the star formation
in the ONC was continuous with the most massive stars starting to be formed no
more than $2\,\myr$ ago. Setting $T_\mathrm{ex}$ in our models to either $0.5$
or $0.7\,\myr$ appears to have only a marginal impact on the final state
of the cluster. Hence, we assume that a more complicated temporal prescription
of the gas removal would not affect our results considerably.

\subsection{The runaway-mass star or black hole}
\label{sec:runaway}
Under the assumption of an invariant canonical IMF and a small initial cluster
radius, the formation of the runaway-mass object via stellar collisions appears
to be an inevitable process
that, together with the high-velocity ejections, decreases the number of OB
stars in a dense star cluster. It's possible detection would definitely
strongly support the scenario of the ONC history presented in this paper.

The current state of the runaway-mass star is, however, not clear. The massive
star would definitely have undergone a fast internal evolution. While successive
merging events may lead to very fast mass growth, stellar winds act in the
opposite way. The state of the runaway-mass object after $\approx2\,\myr$ of
evolution depends on the (dis)balance of these two processes. In the literature,
quite different predictions on this subject can be found. \cite{glebbeek09}
suggest that winds should work in a self-regulatory manner, limiting the maximum
mass and, consequently the star's lifetime. Other authors
\citep[e.g.][]{suzuki07,pvh11} state that stellar winds of the runaway-mass star
will not be able to terminate its growth and it may then
collapse to a massive black hole. There is no observational evidence
for a star with a mass above $50\msun$ in the ONC which could be interpreted as
a runaway-mass object. The most massive stellar member of the ONC,
$\Theta^1$C, is a binary with a total mass of about $45\msun$ and mass ratio
$\approx0.25$ \citep{kraus09}, i.e. the primary has mass $\lesssim 35\,\msun$.
It does not exhibit a
stellar wind strong enough to reduce its mass by several tens of $\msun$.
Hence, our scenario assumes a low efficiency of stellar winds, i.e. continuous
growth of the runaway-mass star which forms a massive black hole at the end of
its lifetime. There is no general consensus
whether the collapse of a star more massive than $150\msun$ will be followed
by an ejection of most of its mass. As there is no evidence for a supernova
remnant in the ONC, we assume most of the mass of the runaway-mass star, if it is
present, to have collapsed directly into a black hole.

The putative black hole in the ONC would be detectable only through
an interaction with its environment. One possibility is an accretion of
surrounding gas which could be detectable
through X-ray radiation. The main source of the gas in the Trapezium region
are stellar winds from the remaining OB stars. We estimate that they produce
at most $10^{-5}\msun\,\mathrm{yr}^{-1}$ within the central $0.2\pc$. However,
the black hole is likely to capture only the gas that
falls within its Bondi radius $\approx100\,\mathrm{AU}$, i.e. the
accretion rate could be $\lesssim 10^{-15}\msun \mathrm{yr}^{-1}$. Such a
highly sub-Eddington accretion will not be detectable in the region confused
with several X-ray young stellar sources.

\begin{figure}[t]
\begin{center}
\includegraphics[width=\columnwidth]{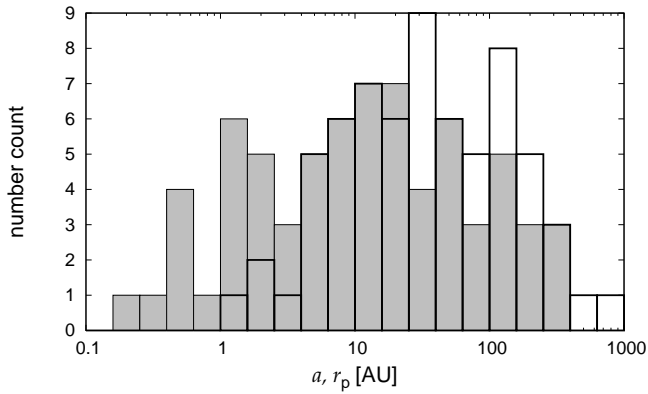}
\end{center}
\caption{Histogram of the semi-major axes ($a$, empty boxes) and pericentre
distances ($r_\mathrm{p}$, full boxes) of the binaries involving the runaway-mass
star in the canonical model. A total of 100 runs were examined, out of which
70 harbour the runaway-mass star with a stellar companion (three of them with
$a > 1000\au$).}
\label{fig:a_dist}
\end{figure}
A better chance for detecting the black hole would be given if it is a member
of a binary. This indeed appears to be the case for about two thirds of the
realisations of the canonical model. However, the typical separation of the
black hole and the secondary appears to be from several tens to hundreds of
$\mathrm{AU}$ (see Fig.~\ref{fig:a_dist}). Even with the relatively large
eccentricites achieved, in no case does the secondary star fill its Roche radius
at the pericentre of its orbit, i.e. we do not expect the black hole to be
a member of a mass transferring system. Still there would be a chance for a
detectable accretion provided the secondary star is massive ($M_\star \gtrsim
10\,\msun$).
In such a case, its stellar wind may produce short periods of enhanced activity
during pericentre passages, but the corresponding luminosity is rather uncertain.
Let us consider the well kown Cyg-X1 system as a kind of template. Having
a black hole mass of $\approx10\,\msun$ and separation from the secondary star
of $\approx0.2\,\au$, it has a luminosity $\approx 2\times10^{37}
\mathrm{erg\,s}^{-1}$ \citep{wilms06} which corresponds to $\approx10^{-4}$ of
the rest-mass energy of the winds emitted by the secondary star ($\approx
10^{-6}\,\msun\,\mathrm{yr}^{-1}$). Assuming the same effectivity in conversion
of the stellar winds into radiation, we estimate the luminosity to be similar to
the Cyg-X1 system, i.e. $\approx10^4\,L_\sun$, for separations
of the order of $1\,\au$. Considering the density of the stellar wind to
decrease with the sqaure of the distance, we obtain a rough estimate of the peak
luminosity $L\approx 10^4 (r_\mathrm{p} / 1\,\au)^{-2} L_\sun$. Acording to
the mass of the putative black hole, the maximum of the emitted radiation is
expected to lie in the X-ray band.
The accretion events should be recurrent with a period ranging from years to
hundreds of years. A shorter period generally implies a smaller $r_\mathrm{p}$,
i.e. larger values of the peak luminosity.
%

\subsection{Velocity dispersion}
Another way of an indirect detection of the massive black
hole in the ONC lies in velocity dispersion measurements. It has been
stated by several authors that the core of the ONC ($r \lesssim 0.25\,\pc$) is
dynamically hot with a velocity dispersion $\gtrsim 4\,\mathrm{km\,s}^{-1}$
\citep[e.g.][]{jw88,vA88,tobin09}. The missing stellar mass required for
virial equilibrum in the innermost region of the ONC was estimated to be
$\gtrsim 2000\,\msun$ which exceeds the observed mass $\approx200\,\msun$
by an order of magnitude. \cite{kroupa01} demonstrate that the globally
super-virial velocity dispersion is readily obtained if the ONC is expanding
now after expulsion of its residual gas, but some of the missing mass in the
inner region could be attributed to the invisible remnant of the runaway-mass
star.

\begin{figure}[t]
\begin{center}
\includegraphics[width=\columnwidth]{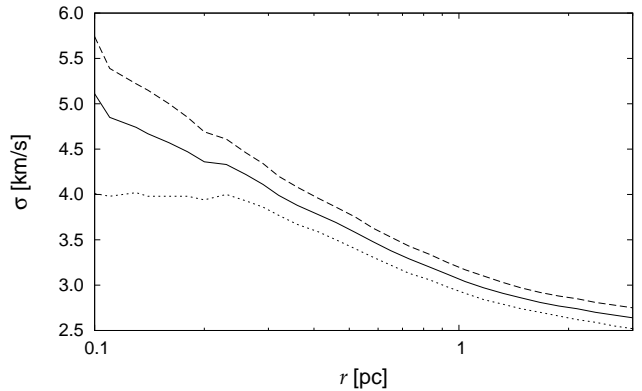}
\end{center}
\caption{Velocity dispersion $\sigma$ within a specified radius $r$ at the
final state of the canonical model. For distinguished bound multiple systems,
the centre-of-mass velocity is counted instead of the proper velocities of
the individual components. Solid line represents the average over all
realisations; thin dashed line corresponds to half of the computations yielding
more massive ($\mbh \gtrsim 120\,\msun$) runaway-mass object, while the dotted
line stands for the computaions with lower $\mbh$.}
\label{fig:dispersion}
\end{figure}
As we show in Fig.~\ref{fig:dispersion}, our canonical model gives a velocity
dispersion $\gtrsim4\mathrm{km\,s}^{-1}$ in the inner $0.25\pc$, which is in
accord with the observational data. In order to inspect the influence of the
mass of the runaway-mass object on the velocity dispersion, we have divided the
100 realisations of the canonical model into two groups of 50 according to the
mass of the most massive object that remained in the cluster. It appears that
the clusters with $\mbh \gtrsim 120\msun$ have a velocity dispersion within
$0.25\pc$ somewhat larger than those with $\mbh \lesssim 120\msun$, nevertheless,
even within the latter group we have $\sigma \gtrsim 3.5\mathrm{km\,s}^{-1}$,
i.e. an apparently super-virial core region. Together with the fact that none
of our numerical experiments led to $\mbh \gtrsim 2000\msun$, virtually required
by the condition of virial equilibrium, this indicates that the large velocity
dispersion can be only partly due to the hidden mass of the runaway-mass object.
Detailed analysis of our models shows that there are two other reasons for
having large velocity dispersion in the core. First, according to our models,
the ONC is a post-core-collapse cluster with centrally peaked density. This
leads to a deep gravitational potential well which allows the central velocity
dispersion to be larger than what would be expected e.g. for the Plummer
profile. Second, in our models and perhaps also in the real ONC, there are likely
to be present several not identified wide binaries with orbital velocities
exceeding $10\,\kms$ which affect the velocity dispersion.

Fig.~\ref{fig:dispersion} indicates that the presence of the dark massive body
influences remarkably (i.e. by more than $1\,\kms$) the
velocity dispersion in a small region of a radius of $\approx 0.1\,\pc$. This is
in accord with an analytical estimate of the influence radius of the black
hole, $r_\mathrm{h} = G\mbh/\sigma^2$, which for $\mbh=100\,\msun$ and
$\sigma=3\,\kms$ gives $r_\mathrm{h} \approx 0.05\,\pc$. Due to the small
number count of stars in such a small region, standard statistical methods do
not represent a robust tool for determining the presence of a dark massive
body. Nevertheless, having this caution in mind, let us discuss the velocity
dispersion of the compact core of the ONC, the four Trapezium stars. According
to the published observational data (see Table~\ref{tab:trapezium}), the
line-of-sight velocity dispersion of the Trapezium is $\sigma_\mathrm{los}
\approx 4.6\,\kms$ which implies a 3D velocity dispersion
$\sigma \approx 7.9\,\kms$. Considering a sphere of radius $r_\mathrm{T}
\approx0.025\,\pc$ covering the four Trapezium stars, we obtain an
estimate of the enclosed mass required for the system to be in virial
equilibrium: $M_\mathrm{bind} \approx r_\mathrm{T}\, \sigma^2 / G \approx
350\,\msun$. Taking into account the stellar mass of the Trapezium
($\approx90\,\msun$), we find it to be an apparently dynamically very hot system.
Our models suggest that the virial equilibrium state can be achieved by
the presence of the runaway-mass object.
\begin{table}[t]
\begin{center}
\caption{Trapezium stars data}
\label{tab:trapezium}
\begin{tabular}{ccc}
name & $M/\msun$ & $v_\mathrm{los} / \kms$ \\
\hline
$\Theta^1A$ & $18.9$ & $33.4 \pm 2$\rule{0em}{2.5ex} \\
$\Theta^1B$ & $7.2$ & $24.0 \pm 2$ \\
$\Theta^1C$ & $44\pm7$ & $23.6$ \\
$\Theta^1D$ & $16.6$ & $32.4 \pm 1$
\end{tabular}
\end{center}
Stellar masses of $\Theta^1A$, $\Theta^1B$ and $\Theta^1D$ are taken from
\cite{hillenbrand97}. Radial velocity data of $\Theta^1A$ and $\Theta^1B$ come
from the C.D.S. -- SIMBAD database; data of $\Theta^1D$ follows
\cite{vitrichenko02}. Both mass and radial velocity of $\Theta^1C$ are
according to \cite{kraus09}.
\end{table}

Yet another, and statistically better founded, argument for a hidden massive
body comes from comparison of the kinematical state of the Trapezium and similar
subsystems found in our numerial models. We have implemented an algorithm
similar to that used by \cite{AG11} for detection of Trapezium-like systems. In
particular, for each OB star we determined its three nearest OB neighbours
(either single OB stars, or bound systems containing at least one OB star).
The set of stars was considered Trapezium-like if all members had the same OB
neighbours. In order to obtain a statistically significant
sample, we examined the last 20 snapshots (covering the
period from $\approx 1.5\,\myr$ to $2.5\,\myr$) of 100 independet realisations
of the canonical model. We found Trapezium-like systems in a third of
cluster snapshots. The mean value of the velocity dispersion of these
systems was found to
be $\approx 7.5\,\kms$. The runaway-mass star was excluded from the search
algorithm. Furthermore, we have distinguished systems according to their
centre of mass distance to the runaway-mass object. In approximately two
thirds of the Trapezium-like systems, the runaway-mass object was found
within a distance smaller than $\max(\Delta r_{ij})$, with $\Delta r_{ij}$\
denoting separations of individual members. The mean velocity dispersion of the
Trapezium-like system of this subset was $\approx 9.0\,\kms$, while it was
only $\approx 4.3\,\kms$ for the remainig cases.

Finally, let us remark that kinematical evidence of a massive black hole in
the ONC may also lie in the $\approx70\%$ probability that it is a member of a
binary (see Sec.~\ref{sec:runaway}). The typical velocity of the secondary star
should be $>10\,\mathrm{km\,s}^{-1}$, i.e. considerably exceeding the observed
central velocity dispersion.

\section{Conclusions}
We have carried out extensive modelling of the dynamical evolution of compact
young star clusters, aimig to reconstruct the history of the ONC. Assuming that
the ONC underwent primordial gas expulsion, we have shown that the ONC must
have been several times more compact than it is now in agreement with
\cite{kroupa00} and \cite{kroupa01}. This implies that its
two-body relaxation time was considerably shorter in the past and,
consequently, that close few-body interactions between massive stars were
rather frequent. We have concentrated on two significant tracks of such
interactions that lead either to high-velocity ejections of massive stars from
the cluster or to their physical collisions that are likely to lead to the formation
of a massive `runaway' merging object. Both of these processes decrease the
number of massive stars in the cluster. Hence, the observed significant lack of
massive OB stars in the ONC further supports our assumption that this star
cluster has undergone a relaxation dominated period since its birth.

We have shown that it is not unfeasible
that the centre of the ONC harbours a black hole of mass $\gtrsim100\msun$ as
the remnant of the massive runaway-mass star. Its presence could be revealed
either by episodic accretion events or by a thorough kinematical study of the
innermost $\approx 0.05\,\pc$ region of the ONC, where the hidden mass would
increase the velocity dispersion above $5 \mathrm{km\,s}^{-1}$. In particular,
we have shown that the observed velocity dispersion
of the Trapezium system can be achieved by the presence of an object of mass
$\approx 150\,\msun$ which is fully consistent with the hypothesis of the
runaway-mass object. Our model also shows that the apparent
super-viriality of the central $\approx 0.25\pc$ can be explained as
a natural attribute of a post-core-collapse dynamical state of a star cluster
with a considerable binary fraction.

The possible detection of the remnant of the merger object could bring new
light into several fields of contemporary astrophysics. First, it would
confirm the hypothesis that star clusters similar to the ONC are being formed
very compact with a half-mass radius of the order of a few tenths of a parsec
\citep{marks12}.
Second, and probably more importantly, it would have important implications for
the evolution of very massive stars and merger products for which we have only
a limited understanding now.

\begin{acknowledgments}
\section*{Acknowledgments}
We thank the anonymous referee for helpful comments.
This work was supported by the Czech Science Foundation via grant GACR-205/07/0052
and from the Research Program MSM0021620860 of the Czech Ministry of Education.
HB acknowledges support from the German Science Foundation through a Heisenberg
Fellowship and from the Australian Research Council through a  Future Fellowship
grant FT0991052.
\end{acknowledgments}

\bibliographystyle{apj}

\begin{appendix}
\section{Theoretical predictions}
Unlike two-body relaxation, interactions among three or more stars allow
considerable energy transfer from one component to another. Therefore,
multiple-body scattering is essential for both processes (merging and
ejections) that lead to the reduction of the number of massive stars
in the cluster. In spite of the chaotic nature of the dynamical evolution of
the star cluster core, it is possible to derive a raw estimate of the rate of
OB stars ejections and collisions.

Our approximation of three-body scattering follows the work of \cite{ps12}:
Consider an interaction of a binary of mass
$M_\mathrm{B}=M_1+M_2$ and a single star $M_\star$. Large accelerations of the
single star are assumed to occur when it passes around one of the binary
components within the semi-major
axis, $a$, of the binary. The cross section of the interaction is assumed to be
determined by gravitational focusing:
\begin{equation}
 \Sigma(a) \approx \frac{2\pi G M_\mathrm{B}a}{v_\mathrm{c}^2}\;,
\end{equation}
where $v_\mathrm{c}$ is the characteristic stellar velocity in the cluster and
$G$ stands for the gravitational constant.

The energy transfer estimate is based on the approximation that the impacting
star moves along a hyperbolic orbit around $M_1$ perturbed by $M_2$. The
typical perturbing force is
\begin{equation}
 F \approx \frac{G M_2 M_\star}{a^2}
\end{equation}
and it acts along the star's trajectory segment of length $\approx a$.
Hence, the energy transfer to the star is
\begin{equation}
 \Delta E_\star \approx \frac{G M_2 M_\star}{a} \approx
 \frac{G M_\mathrm{B} M_\star}{a}\;.
\end{equation}
Here we for simplicity assume the mass of the binary to be of the same order as
the mass of the secondary, $M_\mathrm{B} \approx M_2$. In the cases when
$\Delta E_\star$ is much larger or at least comparable to
the star's energy before the interaction, it will be accelerated to
\begin{equation}
 v_\mathrm{acc} \approx \sqrt{\frac{2G M_\mathrm{B}}{a}}\;.
\label{eq:v_eject}
\end{equation}

Finally, let us assume the distributions of the binary mass, semi-major axis
and eccentricity in a simple power-law form,
\begin{equation}
 n(a, e, M_\mathrm{B}) = 2A\, e\, a^{-1}\, M_\mathrm{B}^{-\alpha}
\label{eq:bindist}
\end{equation}
in the interval $a\in\langle a_\mathrm{min}, a_\mathrm{max} \rangle$,
$e \in \langle 0, 1 \rangle$ and $M_\mathrm{B} \in \langle M_\mathrm{min},
M_\mathrm{max} \rangle$, i.e. $n(a,e,M_\mathrm{B})\,\rd a\,\rd e\,\rd M_\mathrm{B}$
is the number of binaries with semi-major axis, eccentricity and mass in
$\langle a, a+\rd a \rangle$, $\langle e, e+\rd e \rangle$ and $\langle M_\mathrm{B},
M_\mathrm{B}+\rd M_\mathrm{B} \rangle$ respectively. The normalisation constant
then reads:
\begin{equation}
A = \,\biggr\{
\begin{array}{lcl}
 \!\ln^{-1}\!\left( \frac{a_\mathrm{max}}{a_\mathrm{min}} \right)
 \ln^{-1}\!\left( \frac{M_\mathrm{max}}{M_\mathrm{min}} \right)&
 \makebox[2em]{for} & \alpha=1\;, \\
 \rule{0em}{3.5ex}
 \!\ln^{-1}\!\left( \frac{a_\mathrm{max}}{a_\mathrm{min}} \right)
 \frac{1 - \alpha}{M_\mathrm{max}^{1-\alpha} - M_\mathrm{min}^{1-\alpha}}&
 \makebox[2em]{for} & \alpha \neq 1\;. \\
\end{array}
\biggl.
\end{equation}

\subsection*{High-velocity ejections}
The frequency of scattering events of a star with velocity $v_\mathrm{c}$
on binaries with number density $n_\mathrm{B}$ and semi-major axis $a$ is
\begin{equation}
 \nu = \Sigma\,n_\mathrm{B}\,v_\mathrm{c} = \frac{2\pi G M_\mathrm{B}
 n_\mathrm{B} a}{v_\mathrm{c}}\;,
\label{eq:nu_eject}
\end{equation}
where we assumed the binary cross-section to be determined by the gravitational
focusing, i.e. $\Sigma \approx 2\pi GM_\mathrm{B}a / v_\mathrm{c}^2$.
The mean frequency of ejections can be obtained via integration of
(\ref{eq:nu_eject}) weighted by the distribution function (\ref{eq:bindist}):
\begin{equation}
 \bar{\nu}_\mathrm{e} = \int_{M_\mathrm{min}}^{M_\mathrm{max}} \rd
 M_\mathrm{B}\int_{a_\mathrm{min}}^{a_\mathrm{lim}} \rd a\,
 \nu(a,M_\mathrm{B})\, n(a, e, M_\mathrm{B})\;.
\label{eq:nu_mean}
\end{equation}
The upper limit of the semi-major axis, $a_\mathrm{lim}$, has to be set such
that only interactions that lead to acceleration above the escape velocity
from the cluster are considered, i.e.
\begin{equation}
 v_\mathrm{acc}(a_\mathrm{lim}) = v_\mathrm{esc} \approx
 \sqrt{\frac{2G M_\mathrm{c}}{r_\mathrm{c}}}\;,
\label{eq:vesc}
\end{equation}
where $r_\mathrm{c}$ is a characteristic radius of the cluster. Combining
(\ref{eq:vesc}) and (\ref{eq:v_eject}) gives $a_\mathrm{lim} \approx
r_\mathrm{c} M_\mathrm{B} / M_\mathrm{c}$. For the canonical model with
initial $M_\mathrm{c} = 5400\msun$ and $r_\mathrm{c} = 0.11 \pc$
and for $5\msun \leq M_\mathrm{B} \leq 100\msun$ we obtain
$a_\mathrm{lim} \gtrsim 100\mathrm{AU}$ which is the upper limit
of the semi-major axis distribution used in our numerical integrations. Hence,
let us for simplicity assume $a_\mathrm{lim} = a_\mathrm{max} = const.$
which yields
%
\begin{equation}
\bar{\nu}_\mathrm{e} = \,\Biggr\{
 \begin{array}{lcl}
  \displaystyle\frac{2\pi G n_\mathrm{B}}{v_\mathrm{c}}\,
  \frac{a_\mathrm{max} - a_\mathrm{min}}{\ln(a_\mathrm{max}/a_\mathrm{min})}\,
  \frac{M_\mathrm{max} - M_\mathrm{min}}{\ln(M_\mathrm{max}/M_\mathrm{min})}
  & \makebox[3em]{for} & \alpha=1\;, \\
  \displaystyle\frac{2\pi G n_\mathrm{B}}{v_\mathrm{c}}\,
  \frac{a_\mathrm{max} - a_\mathrm{min}}{\ln(a_\mathrm{max}/a_\mathrm{min})}\,
  \frac{1 - \alpha}{2 - \alpha}\,
  \frac{M_\mathrm{max}^{2-\alpha} - M_\mathrm{min}^{2-\alpha}}
  {M_\mathrm{max}^{1-\alpha} - M_\mathrm{min}^{1-\alpha}}
  & \makebox[3em]{for} & \alpha\neq 1\;. \rule{0em}{5ex}
 \end{array}
\Biggl.
\end{equation}
Considering an initial OB binary density $n_\mathrm{B} \approx 10^5 \pc^{-3}$
and velocity dispersion $v_\mathrm{c} \approx 10\mathrm{km\,s}^{-1}$,
$a\in \langle 0.1\au,\, 100\au \rangle$ and $M_\mathrm{B} \in \langle 5\msun,\,
100\msun \rangle$ with a Salpeter mass function ($\alpha = 2.35$) we obtain
$\bar{\nu}_\mathrm{e} \approx 0.1 \myr^{-1}$. Hence, the mean time for an
OB star to undergo scattering on a massive binary that leads to its ejection
from the cluster is $\tau_\mathrm{e} \equiv 1 / \bar{\nu}_\mathrm{e}
\approx 10\myr$ for the canonical model.

\subsection*{Stellar collisions}
Unlike in the case of stellar ejections, we assume that stellar collisions
are caused by binary shrinking due to successive three body interactions.
We further assume all the massive binaries to be hard, i.e. their interaction
with the third body leads to the acceleration of the impact star and growth of
the binding energy of the binary. The frequency of the events of scattering of
a star of mass $M_\star$ on the massive binary is $\nu^\prime = \Sigma
v_\mathrm{c} n(M_\star)\rd M_\star$, where $n(M_\star)\rd M_\star$ is the
number of stars of given mass per unit volume. The rate of the energy transfer
from the binary to the stars of mass in
$\langle M_\star, M_\star+\rd M_\star \rangle$ is
\begin{equation}
 \frac{\rd E}{\rd t} \approx \Delta E_\star \nu^\prime =
 \frac{2\pi G^2 M_\mathrm{B}^2 M_\star}{v_\mathrm{c}}\, n(M_\star)\rd M_\star\;.
\end{equation}
The net rate of energy transfer to stars within the whole mass spectrum gives
\begin{equation}
 \frac{\rd E}{\rd t} \approx \int \frac{2\pi G^2 M_\mathrm{B}^2 M_\star}
 {v_\mathrm{c}} n(M_\star) \rd M_\star =
 \frac{2\pi G^2 M_\mathrm{B}^2 \rho_\star}{v_\mathrm{c}} \;.
\end{equation}
The stars collide when the binary separation at pericentre becomes less than
the sum of the stellar radii. In terms of binding energy and with approximation
$R_\star(M_1) + R_\star(M_2) \approx R_\star(M_\mathrm{B})$, the condition
for collision reads:
\begin{equation}
 E_\mathrm{coll} \approx \frac{1-e}{8}
 \frac{G M_\mathrm{B}^2}{R_\star(M_\mathrm{B})}\;.
\end{equation}
The time required to grow the binding energy from the initial value 
$E_0 \approx \frac{1}{8} G M_\mathrm{B}^2 / a_0$ to $E_\mathrm{coll}$ is
\begin{equation}
 t_\mathrm{coll} \approx (E_0 - E_\mathrm{coll}) \left| \frac{\rd E}{\rd t}
 \right|^{-1} \approx \frac{v_\mathrm{c}}{8\pi G \rho_\star} \left(
 \frac{1-e}{R_\star(M_\mathrm{B})} - \frac{1}{a_0} \right)\;.
\end{equation}
Integration of $t_\mathrm{coll}$ over the whole parameter space of binaries
with the distribution function~(\ref{eq:bindist}) gives the characteristic
time of stellar collisions:
\begin{eqnarray}
 \tau_\mathrm{coll} & \approx & \frac{v_\mathrm{c}}{16\pi G \rho_\star}
 \int_{a_\mathrm{min}}^{a_\mathrm{max}} \! \rd a
 \int_{M_\mathrm{min}}^{M_\mathrm{max}} \! \rd M_\mathrm{B}
 \int_0^{1-R_\star(M_\mathrm{B})/a} \! \rd e
 \left( \frac{1-e}{R_\star(M_\mathrm{B})} - \frac{1}{a_0} \right)
 n(a, e, M_\mathrm{B}) \nonumber \\
 & \approx & \frac{v_\mathrm{c}}{16\pi G \rho_\star}
 \int_{a_\mathrm{min}}^{a_\mathrm{max}} \! \rd a
 \int_{M_\mathrm{min}}^{M_\mathrm{max}} \! \rd M_\mathrm{B} \,
 A\,a^{-1} M_\mathrm{B}^{-\alpha}
 \left( \frac{1}{3R_\mathrm{B}} - \frac{1}{a}
 + \frac{R_\mathrm{B}}{a^2} - \frac{R_\mathrm{B}^3}{3a^3}
 \right)\;,
\label{eq:tau_coll_1}
\end{eqnarray}
where we denoted $R_\mathrm{B} \equiv R_\star(M_\mathrm{B})$; the upper limit
on the eccentricity is set such that the binaries are not collisional
initially. Assuming $R_\mathrm{B} \gg a$, we omit the two least significant
terms in~(\ref{eq:tau_coll_1}) and perform the integration over $a$:
\begin{equation}
 \tau_\mathrm{coll} \approx \frac{A v_\mathrm{c}}{16\pi G \rho_\star}
 \int_{M_\mathrm{min}}^{M_\mathrm{max}} \! \rd M_\mathrm{B}\,
 M_\mathrm{B}^{-\alpha}
 \left[ \frac{1}{3R_\mathrm{B}}\ln\left(
 \frac{a_\mathrm{max}}{a_\mathrm{min}} \right) + \frac{1}{a_\mathrm{max}}
 - \frac{1}{a_\mathrm{min}} \right]\; .
\end{equation}
Finally, using the notation $\tilde{M} \equiv M/\msun$ we obtain for $R_\star(M) =
R_\odot\,\tilde{M}^{0.8}$
\begin{equation}
\tau_\mathrm{coll} \approx  \,\Biggr\{
 \begin{array}{lcl}
  \displaystyle \frac{v_\mathrm{c}}{16\pi G \rho_\star R_\odot}
  \left[ \frac{5}{12}\left( \tilde{M}_\mathrm{min}^{-0.8} -
  \tilde{M}_\mathrm{max}^{-0.8} \right) \ln^{-1} \! \left(
  \frac{M_\mathrm{max}}{M_\mathrm{min}} \right) +
  \left( \frac{R_\odot}{a_\mathrm{max}} -
  \frac{R_\odot}{a_\mathrm{min}} \right) \ln^{-1} \! \left(
  \frac{a_\mathrm{max}}{a_\mathrm{min}} \right) \right]
  & \makebox[3em]{for} & \alpha = 1\;, \\
  \displaystyle \frac{v_\mathrm{c}}{16\pi G \rho_\star R_\odot}
  \left[ \frac{1-\alpha}{0.6-3\alpha}
  \frac{\tilde{M}_\mathrm{max}^{0.2-\alpha} - \tilde{M}_\mathrm{min}^{0.2-\alpha}}
  {\tilde{M}_\mathrm{max}^{1-\alpha} - \tilde{M}_\mathrm{min}^{1-\alpha}} +
  \left( \frac{R_\odot}{a_\mathrm{max}} -
  \frac{R_\odot}{a_\mathrm{min}} \right) \ln^{-1} \! \left(
  \frac{a_\mathrm{max}}{a_\mathrm{min}} \right) \right]
  & \makebox[3em]{for} & \alpha\neq 1\;. \rule{0em}{5.5ex}
 \end{array}
\biggl.
\label{eq:tau_coll}
\end{equation}
The canonical model has an initial central density $\rho_\star \approx 2\times
10^6 \msun\,\pc^{-3}$ which gives $\tau_\mathrm{coll} \approx 40\, \mathrm{Myr}$,
i.e. we estimate stellar collisions to be roughly a factor of 4 less efficient
process for OB star removal than three-body scattering. This is somewhat less
than what our numerical experiment shows. This discrepancy can be due to several
reasons. Most important is probably the fact that stellar collisions may occur
sooner due to  perturbations which lead to a strong growth of orbital eccentricity.
Indeed, the numerical experiments show that most of the collisions occur with
$e \gtrsim 0.99$. Hence, the above given derivation can only serve as an order of
magnitude estimate.

Altogether, the three-body interactions are expected to decrease the number of
OB stars in the canonical model of the ONC by a factor of 2 on a time-scale
of $\approx 5\,\myr$ which is in accord with the results of the numerical experiment.
\end{appendix}
\end{document}